\documentclass{aa}

\usepackage{graphicx}

\begin{document}

\title{Implications of Cosmological Gamma-Ray Absorption} 
\subtitle{II. Modification of gamma-ray spectra}
 
 \author{T. M. Kneiske\inst{1}, T. Bretz\inst{1}, K. Mannheim\inst{1}
 \and D. H. Hartmann\inst{2}}

 \institute{Universit\"at W\"urzburg, Am Hubland, 97057 W\"urzburg, Germany
 \and Clemson University, Clemson, SC 29634-0978, USA}
 \offprints{T.M. Kneiske \email{kneiske@astro.uni-wuerzburg.de}}

 \date{Received/Accepted}

 \abstract{Bearing on the model for the time-dependent metagalactic radiation field
developed in the first paper of this series, we compute the
gamma-ray attenuation due to pair production in photon-photon scattering.
Emphasis is on the effects of varying the star formation rate and the fraction of UV
radiation assumed to escape from the star forming regions, the latter being 
important mainly
for high-redshift sources.  Conversely, we investigate how the metagalactic
radiation field
can be measured from the gamma-ray pair creation cutoff as a function of 
redshift, the Fazio-Stecker relation.
For three observed TeV-blazars (Mkn501, Mkn421, H1426+428) we study the 
effects of gamma-ray attenuation on their spectra in detail.
}
\maketitle     

\section{Introduction}
 
High-energy gamma rays traveling through intergalactic space
can produce electron-positron pairs in collisions with low energy photons from
the metagalactic radiation field (MRF) (Nikishov 1962, Goldreich \& Morrison 1964, 
Gould \& Schreder 1966, Jelley 1966).  
Gamma rays of energy above 1~TeV typically interact with infrared photons 
of wavelengths larger than
1~$\mu$m, such as those predominantly emitted by dust-obscured 
galaxies.  
Gamma-rays of energy below 1~TeV interact with 
near-infrared, optical, and ultraviolet photons, mostly from stars.

Following the discovery of extragalactic TeV gamma ray sources (Punch et al. 1992), 
the effects of cosmological pair creation on their spectra have been studied by a number 
of authors (e.g., Malkan \& Stecker 1998, Konopelko et al. 1999, Primack et al. 1999). 
Results differ by a large margin, owing to the different models for the MRF employed by
the authors, and lead to extreme physical interpretations of the observed gamma-ray spectra.
The debate culminates in the claim of Meyer \& Protheroe (2000) that the weakness of the
observed gamma-ray attenuation might have to be remedied by a threshold anomaly for the
pair creation process, such as predicted in certain (ad hoc) models of quantum-gravity which
violate Lorentz invariance
(Stecker \& Glashow 2001).

Discrepancies between the models for the metagalactic radiation field can be 
traced to
the different formalisms which are used (for a detailed discussion see Hauser \& Dwek 2001). 
The simplest method (backward evolution) is to extrapolate 
present day data or template spectra to high redshift in a certain wavelength range 
(Malkan \& Stecker 2001). Cosmic chemical evolution models
self-consistently describe the temporal history of globally averaged 
properties of the Universe (Pei, Fall \& Hauser 1999). 
Semi-analytical models are invoking specific hierarchical structure 
formation scenarios to predict the MRF.
In our approach we developed (Kneiske et al. 2002; Paper I) a semi-empirical, forward-evolution 
model for the optical-to-ultraviolet MRF and for the infrared part a
backward evolution model based on the data obtained from recent
deep galaxy surveys. The model parameterizes the main observational 
uncertainties, (i) the
redshift dependence of the cosmic star formation rate, and (ii) the fraction of UV radiation
released from the star forming regions.

The expected effects of gamma ray absorption vanish below 10 GeV 
(out to redshifts of $z\sim 200$, see
Zdziarski \& Svensson 1989).  On the contrary, at energies above 300 GeV 
gamma rays suffer absorption for sources at redshifts $z>0.2$. This is in line with
current observations.  The satellite-borne EGRET-All-Sky-Survey has resulted in the 3rd EGRET
catalog (Hartman 1999)
of 93 blazars with gamma ray emission between 100 MeV and 10 GeV.
The ground-based pointing telescopes 
Whipple, HEGRA, CANGAROO, and CAT have found only 6 well established blazars 
at energies above 300 GeV searching for gamma ray emission from cataloged
sources, in spite of their superior sensitivity.   The 
few detected sources indeed have very low redshifts ($z<0.2$).  Intrinsic absorption in the
gamma ray sources, e.g. due to the strong infrared radiation field
associated with a dust torus, could give rise to a redshift-independent pair creation cutoff
(Donea \& Protheroe 2003).

Only three blazars Mkn 421, Mkn 501, and H1426+428 were bright enough to determine their spectra.
Mkn 501 and Mkn 421 are showing a turn-over at almost the same energy 
adopting a power-law-times-exponential model.
The spectra show some small changes with flux, and it is under investigation whether this
affects the cutoff energy or not.  Both sources have almost the same redshift $z=0.03$
and should show cosmological absorption at the same energy of $6-18$~TeV.
In spite of large systematic errors, several attempts to probe galaxy evolution from the  
column depth of the extragalactic infrared photons inferred from the gamma ray data have been launched
(Stecker \& de Jager 1993, Biller et al. 1995, Madau  \& Phinney 1996,  Primack et al. 1999, 
Renault et al.~2001, Kneiske et al. 2002).
A third source at a redshift of $z=0.129$ has very recently been discovered. The statistics
of the signal are poor, probably resulting from absorption much stronger at four times the
distance than for the other blazars
(Aharonian et al. 2002, Petry et al. 2002, Costamante et al 2003).

If gamma-ray sources could be detected at redshifts $z>0.2$ using imaging air
Cherenkov telescopes with threshold energies as low as 10 GeV, such as the MAGIC telescope,
one could infer indirectly the MRF from infrared to ultraviolet wavelengths.

The plan of this paper is to employ various parameter sets for the calculation of the MRF and
to study their effects on gamma-ray absorption.  Friedman-cosmology parameters were fixed to
the values $\Omega_{\rm m}=0.3$, $\Omega_\Lambda=0.7$, and $h=0.65$ 
corresponding to the $\Lambda$CDM cosmology.
Five generic constellations of MRF model parameters
which lead to MRF spectra consistent with observations, and which bracket the range of
allowed values,
are described in Sect.~2. 
In Sect.~3 we will use the optical depth of gamma-rays to calculate the intrinsic spectra 
of Mkn501, Mkn421 and H1426+428 from the observed spectra. The absorption at higher redshifts
will be discussed in the fourth section introducing the Fazio-Stecker relation.

   \begin{figure*}
   \centering
   \includegraphics[width=12cm]{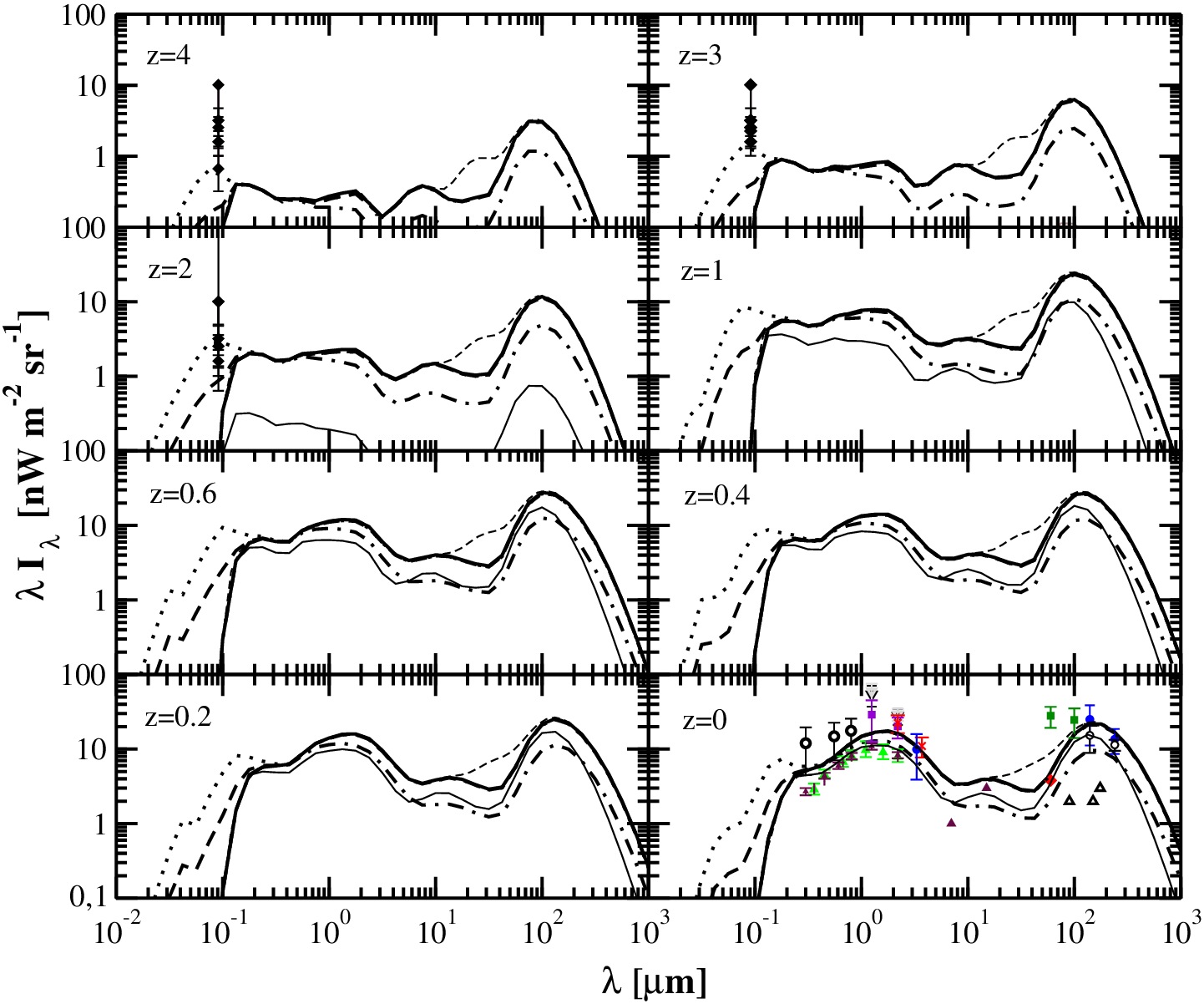}                
   \caption{Comoving-frame metagalactic radiation field (including UV component) at various redshifts.
   "Best-fit" model, thick solid line; "
   Warm-Dust" model, thin dashed line ;"Low-IR" model, dot-dashed line; "Low-SFR" model, 
    thin solid line;
   "Stellar-UV" model, dashed line; and "High-stellar-UV" model, dotted line.
   Data at $z=2,3,4$ are taken from Scott et al.~(2000); data at $z=0$, see Paper I.} 
   \label{fig1}
   \end{figure*}

 \section{The Metagalactic Radiation Field (MRF)}

The model developed in Paper I accounts for the emission from stars, ISM, and dust in galaxies.
Since galaxies strongly obscured by dust do not show up in optical galaxy surveys, recent infrared 
and sub-millimeter surveys were additionally taken into account.  The latter surveys show that more than
half of the cosmic star formation might be hidden in obscured galaxies (Chary \& Elbaz 2001).

We will briefly outline the main ideas behind the MRF model developed in
Paper I (Kneiske et al.~2002).
The power spectrum of the MRF in the comoving frame is given by
 \begin{equation} P_\nu(z) = \nu
 I_{\nu}(z) = \nu \frac{c}{4\pi} \int_z^{z_m}  \mathcal{E}_{\nu'}(z')
\left | \frac{dt'}{dz'} \right |  dz' ~ , \label{eq:hinter} \end{equation}
 with $\nu'=\nu(1+z')/(1+z)$
 and

 \begin{equation}
 \mathcal{E}_{\nu}(z) = \int_z^{z_{m}}
 L_{\nu}(t(z)-t(z'))\dot{\rho}_{\ast}(z') \left | \frac{dt'}{dz'} \right |
 dz' ~,
 \label{eq:emislambda}
 \end{equation}

For the spectra $L_{\nu}(\tau)$ of a simple stellar population as a function 
of age  $\tau$ we used
population synthesis models by Bruzual \& Charlot (1999) with low metalicity.
The absorption due to interstellar medium is modelled by uniform distributed
dust and gas.
In Paper I, the gas around the stars is assumed to absorb all UV photons originating from young,
massive stars. Here we introduce a new parameter which is the fraction 
of the ionizing photons that escapes from a galaxy, e.g. through super-bubbles blown into the 
ISM by supernovae (cf. Stecker et al. 1998). This fraction is, depending on the galaxy type, 
quite low (i.e. Starbursts: $f_{esc}\leq$ 6\% Heckmann et al. 2001).
Note that UV photons from active galactic nuclei (AGN), which might contribute considerably
at UV wavelengths, are not considered at this stage.  
We adopt an average extinction curve for the dust absorption.
The reemission is calculated as the sum of three modified Planck spectra
\begin{equation}
 L_{\lambda}^d(L_{bol}) = \sum_{i=1}^{3} c_i(L_{bol})\cdot Q_\lambda \cdot B_\lambda(T_i)
 \label{eq:planck} \end{equation}
where
$Q_\lambda \propto \lambda^{-1}$ and $L_{bol} = L_{bol}(\tau)$, here $\tau$ is the
age of the stellar population.
The three components characterize cold ($c_1$, $T_1$) and warm ($c_2$, $T_2$) dust and the contribution due to
PAH molecules ($c_3$, $T_3$).
To get the best-fit values for the parameters we used a sample of galaxies
detected with IRAS at 12$\mu$m, 25$\mu$m, 60$\mu$m and 100$\mu$m.
The warm dust component has its maximum around 50$\mu$ and is quite
low in our best-fit model.Increasing $c_2$, we
obtain a model with an enhanced fraction of warm dust, the Warm-Dust model.

The global star formation rate (SFR) $\dot{\rho_\ast}(z)$ consists of two 
components.

\begin{equation} 
 \dot{\rho_\ast}(z) = \dot{\rho_\ast}^{\mathrm{OPT}}(z) + \dot{\rho_\ast}^{\mathrm{ULIG}}(z) 
\end{equation}
The first component accounts for the stars which can be optically detected, 
the other coming from stars hidden by dust which can only be seen
looking at infrared or submm wavelength.
Each of the SFRs can be approximated with a simple
 broken power law
 \begin{equation} \dot{\rho_\ast}(z) \propto (1+z)^\alpha \end{equation}
 with $\alpha=\alpha_m>0$ for $z \le z_{peak}$ 
 and $\alpha=\beta_m<0$ for $z >z_{peak}$. So each SFR 
 provides four parameters $\alpha_m,\ \beta_m,\ z_{\mathrm{peak}}\ \mathrm{ and }\
\dot{\rho_\ast}(z_{\mathrm{peak}})$. The values we used are shown
in Table \ref{tab1}.

\begin{figure}
\centering
   \includegraphics[width=8cm]{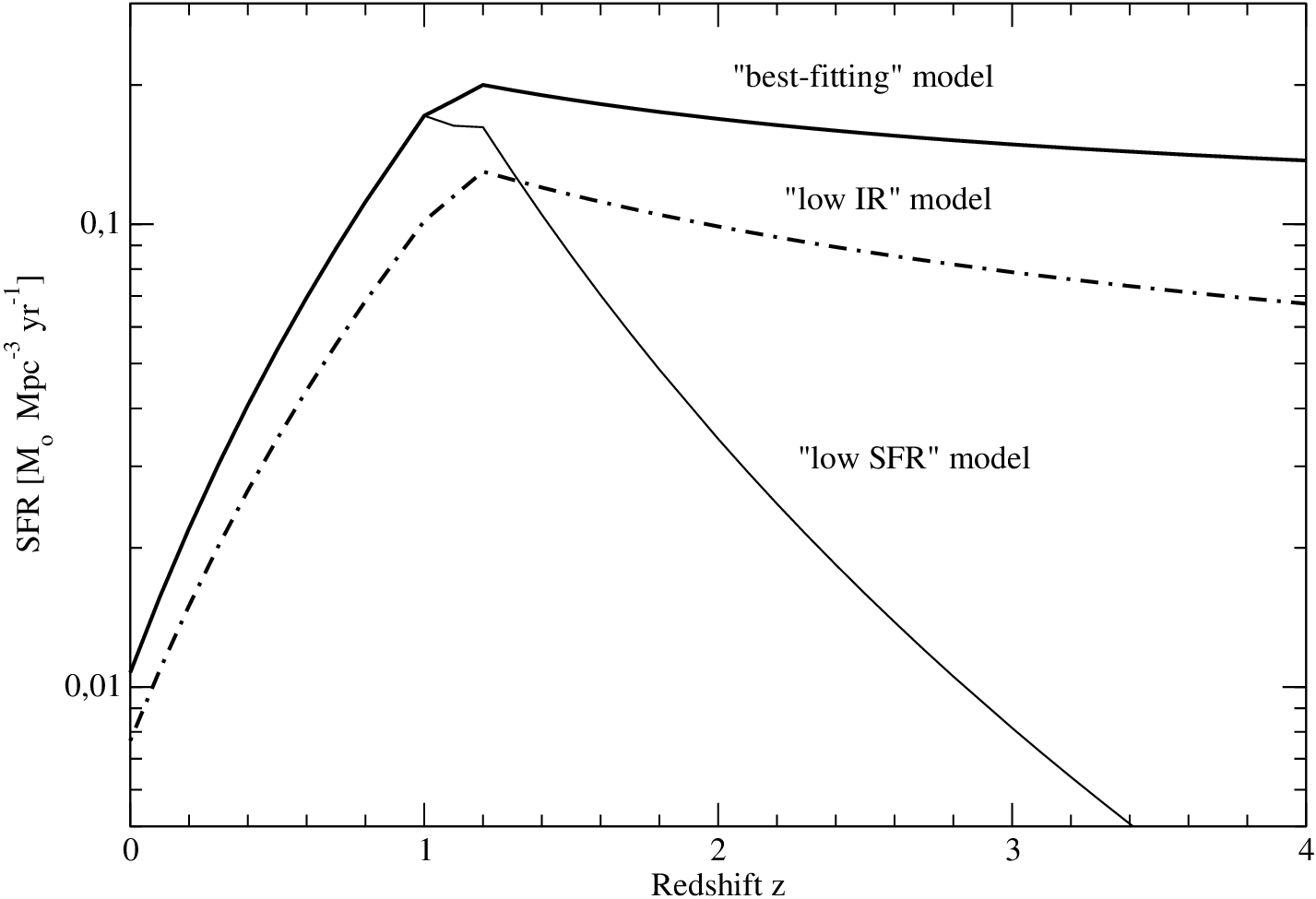}
   \caption{Total global star formation rate for different models SFR = SFR$_{OPT}$ + SFR$_{LIG}$
    (for details, see Paper I)}
   \label{fig0}
   \end{figure}

For reasons outlined in Paper I, it is sufficient to consider 
the star formation rates and the UV escape fraction as the dominant parameters.
In the following, we will use six generic parameter constellations leading to MRF predictions bracketing
the data, see Fig.~\ref{fig1}.

\subsection{Best-fit model}

 \begin{table}
\centering
\caption{Parameters (Definitions see Paper I)}
 
 \begin{tabular}{lllll}
 \hline \hline\
 & $\alpha$ & $\beta $ & $z_p$ & $ \dot{\rho_\ast}(z_p)$  \\
 &&&&[$M_\odot$ Mpc$^{-3}$ yr$^{-1}$]        \\
 \hline 

 &&&&\\
 \bf{Best-fit model}  &&&& \\
 $SFR_
 {OPT}$   & 3.5 & -1.2 & 1.2 & 0.1 \\
 $SFR_{LIG}$   & 4.5 & 0    & 1.0 & 0.1 \\
 $f_{esc}=0$ &	   &	  &     &     \\
 $c_2 = 10^{-24}$ &&&&\\

 &&&&\\
 \bf{Warm-Dust model}  &&&& \\
 $SFR_
 {OPT}$   & 3.5 & -1.2 & 1.2 & 0.1 \\
 $SFR_{LIG}$   & 4.5 & 0    & 1.0 & 0.1 \\
 $f_{esc}=0$ &	   &	  &     &     \\
 $c_2 = 10^{-23.4}$ &&&&\\

 &&&&\\
 \bf{Low-IR mode}l  &&&& \\
 $SFR_{OPT}$   & 3.5 & -1.2 & 1.2 & 0.1 \\
 $SFR_{LIG}$   & 4.5 & 0    & 1.0 & 0.03 \\
 $f_{esc}=0$ &	   &	  &     &     \\
 $c_2 = 10^{-24}$ &&&&\\

 &&&&\\
 \bf{Low-SFR model}  &&&& \\
 $SFR_{OPT}$   & 3.5 & -5 & 1.2 & 0.1 \\
 $SFR_{LIG}$   & 4.5 & -5 & 1.0 & 0.1 \\
 $f_{esc}=0$ &	   &	  &     &     \\
 $c_2 = 10^{-24}$ &&&&\\

 &&&&\\
 \bf{Stellar-UV model} &&&& \\
 $SFR_{OPT}$   & 3.5 & -1.2 & 1.2 & 0.1 \\
 $SFR_{LIG}$   & 4.5 & 0    & 1.0 & 0.1 \\
 $f_{esc}=1$ &	   &	  &     &     \\
 $c_2 = 10^{-24}$ &&&&\\

 &&&&\\
 \bf{High-Stellar-UV model} &&&& \\
 $SFR_{OPT}$   & 3.5 & -1.2 & 1.2 & 0.1 \\
 $SFR_{LIG}$   & 4.5 & 0    & 1.0 & 0.1 \\
 $f_{esc}=4$ &	   &	  &     &     \\
 $c_2 = 10^{-24}$ &&&&\\
 &&&&\\
\hline
\end{tabular}
\label{tab1}
\end{table}

The parameters used in the best-fit model interpolate best the data from galaxy number counts at optical and 
infrared wavelengths, and direct measurements of the extragalactic background, i.e. the present-day MRF.
The model MRF shows a sharp cut-off at 0.1 $\mu m$ due to the total absorption of ultraviolet starlight due to 
interstellar gas.

\subsection{Warm-Dust model}
This model is almost identical to the Best-Fit model except for the amount
of warm dust in the interstellar medium. The different dust types are calculated
to fit the line intensities detected in a sample of infrared galaxies by IRAS at
12$\mu$m, 25$\mu$m, 60$\mu$m and 100$\mu$m. The range between 25$\mu$m and
60$\mu$m is not well determined. 
We raise the amount of warm dust at 80~K
to the maximum determined by the spectra of individual infrared galaxies. 
The corresponding change is clearly noticeable in all panels of Fig.~\ref{fig1} and in Fig.~\ref{fig4} for sources 
at redshifts smaller than $z\approx0.02$.

\subsection{Low-IR model} 
The infrared part of the low redshift (z$<$0.5) MRF is important for the absorption of gamma rays from
low redshift sources.  In order to consider the least possible gamma-ray attenuation, we
adopt a low-IR model, where we choose the infrared
star formation rate as low as allowed by observational lower limits on sub-mm
galaxy number counts (e.g. SCUBA $\approx 0.06$ for $2<z<5$ Hughes et al. (1998) 
, see Fig~\ref{fig0}).
For the sake of demonstration we accept that the present-day background 
in this model somewhat drops below
the lower limits from ISO at 15$\mu m$ and from IRAS at 60 $\mu m$.

\subsection{Low-SFR model} 
The cosmic star formation rate at high redshifts is still a matter of debate.
Therefore, we consider a steep decline of the star formation rate at high redshifts
(see Fig~\ref{fig0}) as an extreme to study the effect of high z star formation and  
gamma ray attenuation (as opposed to the plateau in the other models).
The star formation rate in the low-SFR model drops so rapidly that at $z=2$ it is already
an order of magnitude below the star formation rate of the other models.

\subsection{Stellar-UV model} 
This model differs from the best-fit model in the escape fraction parameter which is 
set to unity allowing for the entire stellar UV light to escape into the metagalactic medium. 
Due to the missing reprocessed UV radiation,
the part of the MRF at wavelengths $\lambda > 0.2~\mu$m is not exactly the same
as in the best-fit model (although it might appear so on the double-logarithmic plot
washing out minute details).

\subsection{High-stellar-UV model} 
The proximity effect (Bajtlik, Duncan \& Ostriker, 1988) allows for a measurement
of the metagalactic UV radiation field at a wavelength of 912~\AA, even at very high redshifts. 
Comparing our results with the recent compilation of data
from Scott et al.~(2000) (upper panels Figure~\ref{fig1})
shows that even the model with $f_{esc}=1$ lies below the data. The reason for this
discrepancy could be that the MRF model does not yet include the UV emission produced by AGN.
Haardt \& Madau (1996) found values for an AGN contribution to the UV MRF of
1.64~nW~$\rm m^{-2}~s^{-1}~sr^{-1}$ at a redshift of $z=2.5$
which would account for the value measured with the proximity effect method.
The UV excess over the stellar MRF is about a factor of four and should make a significant 
difference for high redshift gamma-ray absorption. As a zeroth-order approximation, 
we assume an empirical UV background component by multiplying the UV component 
of the stellar-UV model 
by a factor of four, matching at the Lyman break by linear interpolation. Chosing a, perhaps 
more realistic, power law representation of the template would not make a significant difference
for this analysis.

  \begin{figure*}
  \centering
  \includegraphics[width=12cm]{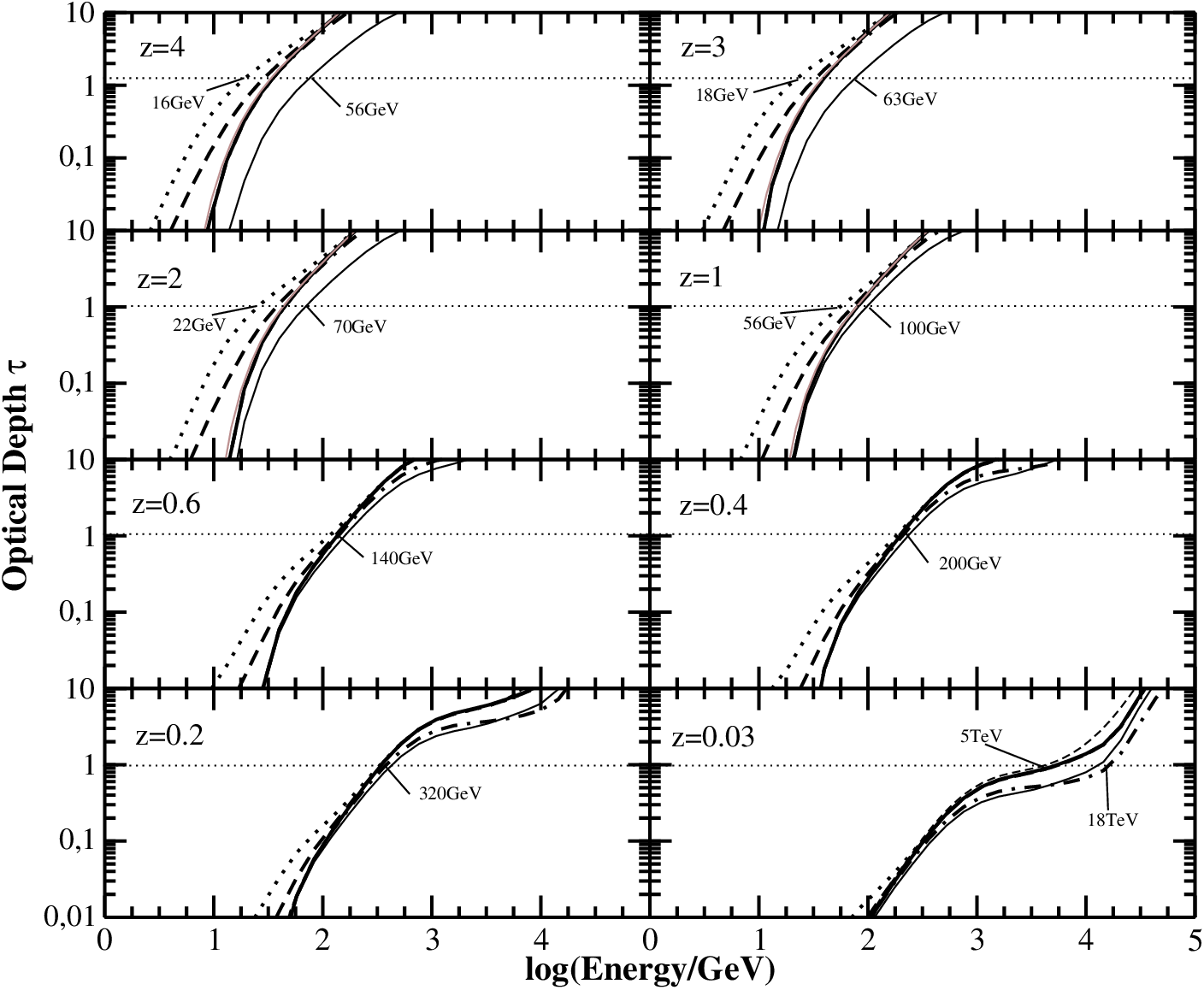}
  \caption{Optical depth for various redshifts adopting a $\Lambda$CDM 
  cosmology.  The labeling of the line styles is the same as in Fig.~1. The crossing point with
  the line $\tau=1$ defines the exponential cutoff energy. }
    \label{fig2}
  \end{figure*}

\subsection{The Optical Depth}

The optical depth for pair creation for a source at redshift $z_q$, and at an observed energy
$E_\gamma$, is obtained from 

\begin{equation}
\tau_{\gamma \gamma}(E_\gamma,z_q) =c\int_{0}^{z_q} \int_{0}^{2} \int_{\epsilon_{gr}}^{\infty} 
\frac{dl}{dz'} \frac{\mu}{2} \cdot {n(z,\epsilon)} 
\cdot \sigma_{\gamma\gamma}(E_\gamma,\epsilon,\mu,z') 
\  d\epsilon\  d\mu\ dz' 
\end{equation}
 with the cosmological line element $\frac{dl}{dz'}$, the angle $\theta$ between the 
interacting photons $\mu = \cos(\theta)$, the number density of the MRF $n(z,\epsilon)$ as
a function of reshift and MRF photon energy and the pair-production cross
section $\sigma_{\gamma \gamma}$.

\begin{figure*}
   \centering
   \includegraphics[width=10cm]{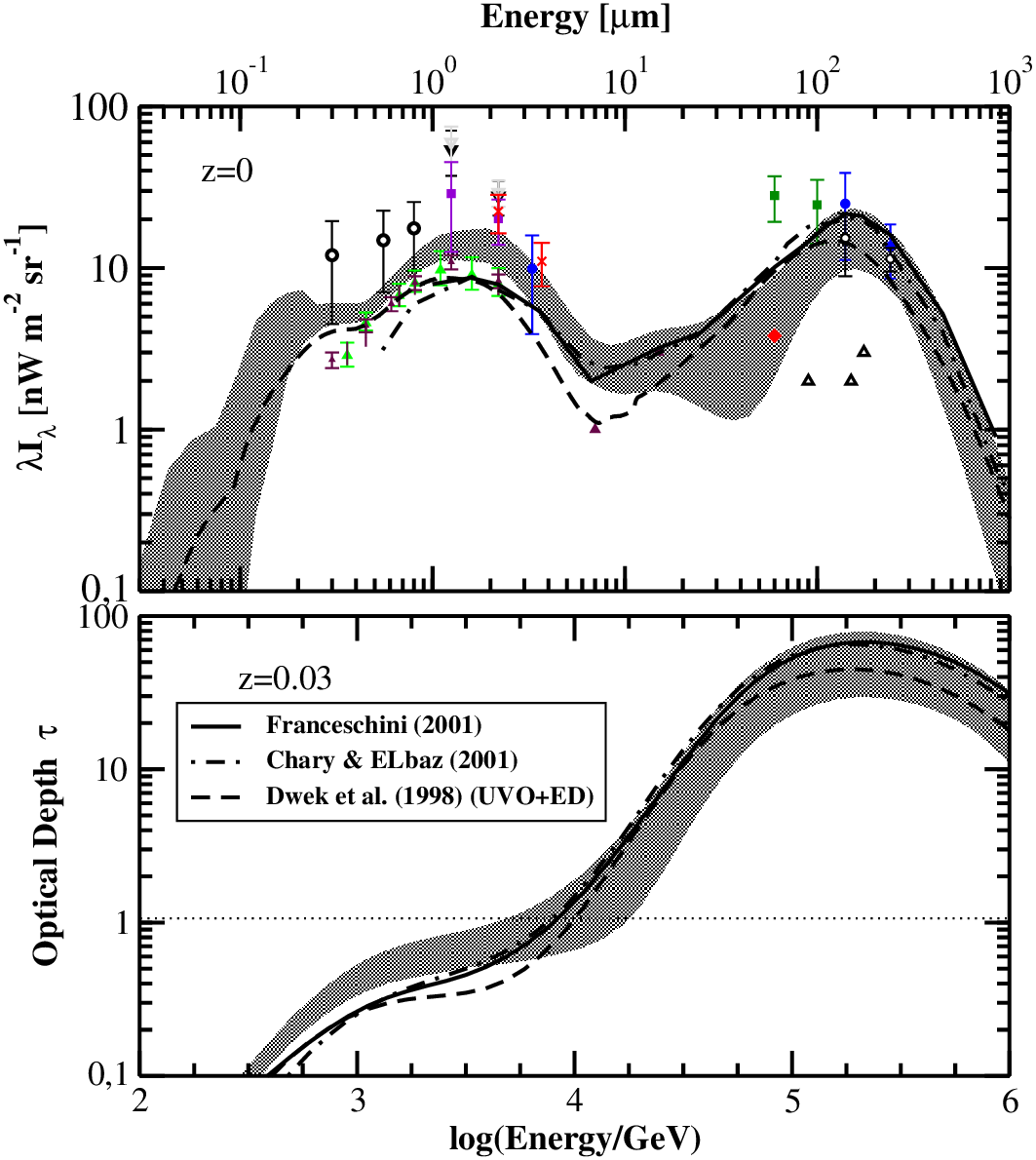}
   \caption{MRF (upper panel) and the optical depth (lower panel) at redshift $z=0$ and $z=0.03$ respectively. Shaded regions denote the regions
   bounded by our models. Dwek et al. (1998), solid line; Franceschini et al. (2001), dashed line;
   Chary\&Elbaz (2001), dot-dashed line. Data references see Fig.~1.}
   \label{fig1a}
   \end{figure*}

By comparing the generic MRF models (see Fig.~\ref{fig2}), it can be seen that the optical depth
from $0.2 < z < 1$ is rather insensitive to the choice of the parameters in the MRF model. 
The interaction mainly takes place with optical MRF photons, which are emitted
by stars and undergo no absorption of the ISM. The rather small differences in
pure stellar models have only a weak effect on the optical depth.
However, at smaller redshifts effects become strong. For example, using the best-fit model we 
obtain a cut-off energy of $\sim5$~TeV, while for the low-IR model the energy is
 $\sim18$~TeV.
Only at very low redshifts ($z<0.1$)
the effect of the Warm-Dust model is noticeable.
The cut-off energy for a source at $z=0.03$ (redshifts $z>0.02$) remains 
largely unaffected, see 
Fig.~\ref{fig2} and Fig.~\ref{fig1a}, since the change is relevant only for
energies above 10~TeV and occurs at an optical depth from 2 to 10.
Only Fig.~\ref{fig3} is showing such high gamma energies at the high end 
of blazar spectra. The upper limit of the shaded region above 10 TeV are 
made by the warm-dust model. Looking at spectra of single sources
is the only method to probe this part of the MRF
(e.g. Biller et al 1995, Stanev \&Franceschini 1998, Guy et al. 2000, Renault et al. 2001).
But there are
two problems, the low statistic of the TeV data and the production processes 
of such high energy photons in AGNs which are still under discussion.

At higher redshifts, the optical depth due to interactions with the UV part of the MRF
become important.  Consequently, the cut-off energy decreases by adding the UV components 
to the MRF model.
By contrast, the cut-off energy increases by lowering the star formation rate at high
redshifts in the low-SFR model. For example, the
 cut-off energy for a source at a redshift of $z=4$ ranges between $\sim16$~GeV for the high-stellar-UV model 
and $\sim40$~GeV for the low-SFR model.

A comparison with other models is shown in Fig~\ref{fig1a}. The range bounded
by our models
contains the results of Dwek et al. (1998), 
Chary\& Elbaz (2001), and Franceshini et al. (2001) in the infrared part of the spectrum.
There are differences in the optical part where the other models come
just close to the lower limits given by the HST data. At low redshifts, this
part is irrelevant for gamma-ray absorption due to the threshold condition, so the
difference is of no concern to us.

\section{Spectral modifications of nearby blazars}

   \begin{figure*}
   \centering
   \includegraphics[width=12cm]{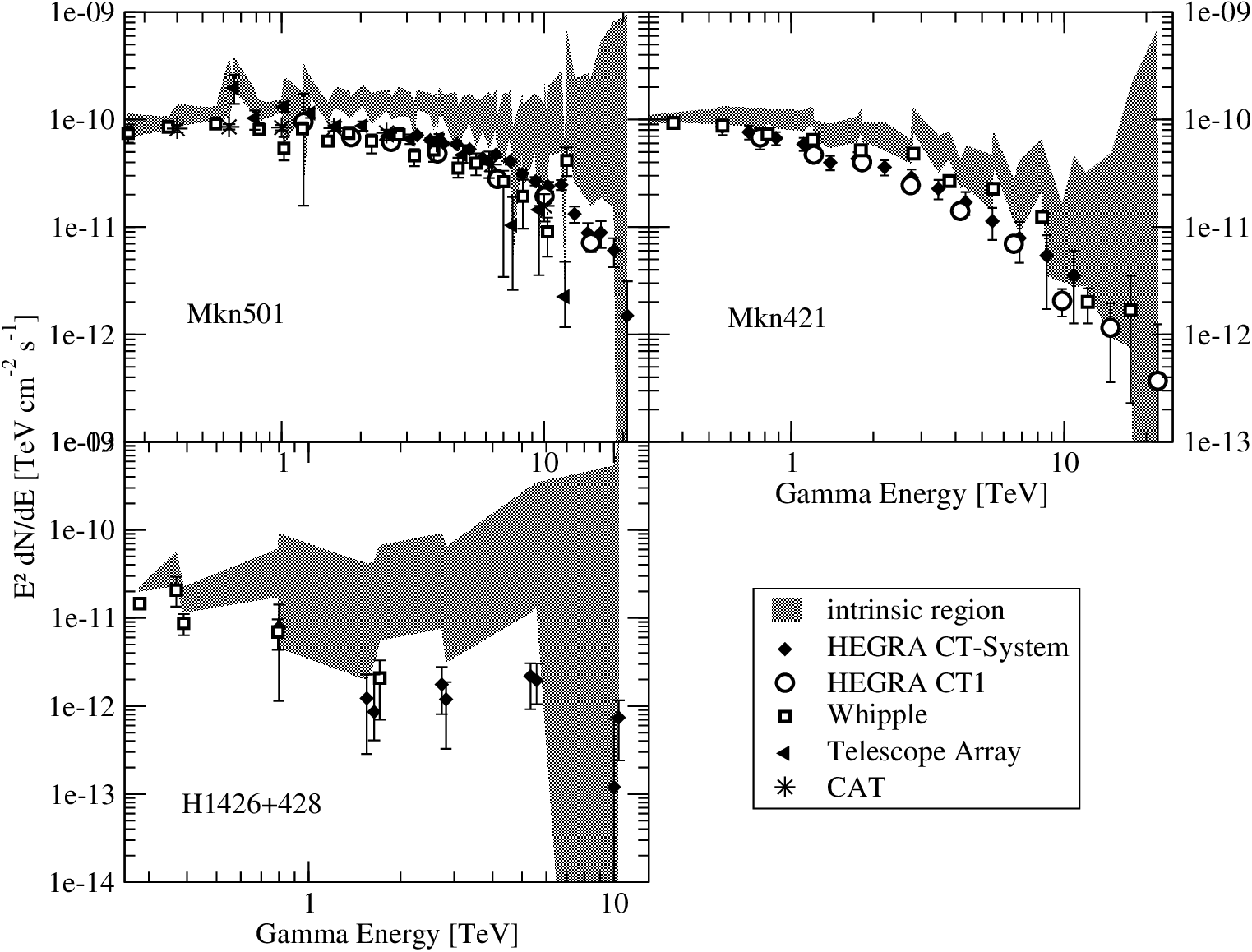}
   \caption{World-data sets for three TeV blazars, and ranges of their 
   intrinsic ("de-absorbed") spectra.}
   \label{fig3}
   \end{figure*}

A number of extragalactic gamma-ray sources have been 
detected with imaging air-Cherenkov telescopes (Table 6, Horan et al. 2002).
Three of them (with redshifts $z=0.03, 0.03, 0.129$)
were bright enough to resolve their spectra in the TeV energy band, as shown in
Figure~\ref{fig3}\footnote{World-data set for Mkn501: 
Aharonian et al. (1999, 2001), Kranich et al. (2001), 
Krennrich et al. (1999), Djannati-Atai et al. (1999) and Hayashida et al. (1998);
for Mkn421: Kohnle et al. (2001), Cortina et al. (2001) 
and Krennrich et al. (2001); for H1416+428: 
Petry et al. (2002), Horan et al. (2002),  
Aharonian et al. (2002) and Costamante et al. (2003)}.

The observed spectra are modified by gamma ray attenuation, i.e. 
\begin{equation}
F_{\rm obs}(E)=F_{\rm int}(E)\exp[-\tau_{\gamma\gamma}(E,z)]
\end{equation}
where $\tau_{\gamma\gamma}(E,z)$ is given by Eq.~1 (see Figure~\ref{fig2} for a number
of examples). 
Note that we neglect secondary gamma rays arising from cascading
in the metagalactic radiation field, which are discussed in a 
separate paper (Bretz et al. 2003).
We used the best-fit model, the Warm-Dust model and the low-IR model to bracket the range
of the unabsorbed (intrinsic) spectra. Using the other generic models 
would not alter the results, 
since the optical depth at low redshifts is practically independent
of the ultraviolet part of the MRF.

The intrinsic spectrum of Mkn501 (power per bandwidth) shows a moderate peak
at 1-4 TeV. In a similar analysis, de Jager \& Stecker (2002) found
5-9 TeV for the peak depending on the infrared model used.
They claim that this peak could correspond to the observed
X-ray peak at 50-100 keV (adopting an SSC model). While Krennrich\&Dwek (2003)
found quite lower values for Mkn501 $785\pm153$GeV to $2390\pm127$GeV.

The intrinsic spectrum of Mkn421 still seems to show a shallow turnover.
Assuming an exponential cutoff, the energy changes from 3~TeV (observed, i.e.
absorbed) to 4~TeV (intrinsic, i.e. unabsorbed). 
A peak is very hard to see, but if there is a maximum it is consistent with the
values found by Krennrich\&Dwek (2003) and would lie between 0.5 and 2 TeV.
This could  
be the signature of the inverse Compton peak reflecting the X-ray peak at 6-8 keV.
To further disentangle observed and intrinsic spectra, it is helpful to look at flux-dependent spectra, using
the defining blazar property of being highly variable sources.
The Mkn421 high-flux spectrum seems to be curved stronger than
the low-flux spectrum which resembles a power law (Aharonian et al. 2002(I)).  
However, the statistical
bias in these studies is non-negligible, and no conclusive evidence has emerged.
A conservative estimate to study the location of the peak at different flux levels
is introduced in Krennrich\&Dwek (2003) and will be discussed in more detail
in a paper by the same authors. They found a shift in the peak energy between the 
lowest and highest flux levels.
The cut-off could also be of an intrinsic origin, if strong radiation fields
surround the gamma-ray production zone (Mannheim 1993, Donea \& Protheroe 2003).
Strong radiation fields in the far-infrared to near-infrared wavelength ranges
result from irradiated dust tori generally surrounding the central engine in AGN.
However, low-redshift blazars (BL Lacertae objects) 
seem devoid of massive tori, and show sub-Eddington accretion implying
weak irradiation.  Moreover, it is at present not clear how deep in the dust torus
the gamma rays are produced rendering this mechanism difficult to estimate.
Theoretical SSC models (Caprini 2002) or proton blazar models (Muecke et al. 2003)
including time variability do not yet predict
the location of gamma ray emission zone relative to the dust torus.

Inferring a spectrum for H1426+428 is difficult owing to poor statistics.  Inspecting
the range of the likely intrinsic spectrum, the low flux seems to be the consequence
of heavy absorption owing to the comparatively high redshift of the source (four times
larger than for Mkn421 and Mkn501, respectively).
The expected exponential cut-off energy at a redshift of $z=0.129$
obtains values of 100-200 GeV, i.e. at energies below the threshold energy of the detecting instruments (Whipple, HEGRA). 

Calculating the intrinsic spectral range we find for the best-fit model the same result
as Costamante et al. (2003). The energy spectrum increases with energy. However,  inspection of
 Figure~\ref{fig3} shows that the intrinsic
energy flux spectrum inferred from the low-IR model remains rather flat, or
even shows a shallow downturn, implying a peak energy an order of magnitude
lower than in the calculation of Costamante et al. (2003). 
As shown in Costamante et al. (2001), the X-Ray peak is at an energy around or
larger than 100~keV, and this would argue in favor of the gamma-ray peak larger than 12~TeV
(adopting an SSC model).  In Bretz et al. (2003), we discuss in detail the fate of the absorbed
gamma ray photons which carry a substantial energy flux.

\section{The Fazio-Stecker relation}


The energy-redshift
relation resulting from the cosmic gamma-ray photosphere 
$\tau_{\gamma\gamma}(E_\gamma,z)=1$
depends on the column-depth of the absorbing
photons, as can be seen from inspection of Eq.~(6).  
We coin this relation, plotted in Fig.~(6), which proves to be very
useful to study the MRF, the ''Fazio-Stecker relation (FSR)'' 
(first shown by Fazio \& Stecker 1970)\footnote[1]{In 1968,
 Greisen has already suggested (in a lecture 
Brandeis Summer Institute in Physics) that pair-production at high redshift 
between optical and gamma photons would produce a cut-off around 10 GeV. }.
The theoretically predicted FSR (depending on the MRF model and cosmological
parameters) can then be compared with a measured one, by determining
e-folding cut-off energies for a large sample of gamma ray sources at
various redshifts.
Two important corollaries follow
from inspecting the Fazio-Stecker relation: (i) gamma-ray telescopes with thresholds much lower than
40~GeV are necessary to determine the cut-off for sources with redshifts around the maximum
of star formation $z\sim 1.5$, and (ii) gamma-ray telescopes with a threshold below 10~GeV have
access to extragalactic sources of any redshift (another cosmological attenuation effect sets
in at $z\sim 200$, Zdziarski 1989).

The main obstacle for this method to indirectly measure the MRF by
achieving convergence between theoretical and observed FSR is
the uncertainty about the true shape of the gamma ray spectra
{\em before} cosmological absorption has ocurred.
In the simplest case, the intrinsic spectra would be just
power law extensions of the (definitively unabsorbed) lower
energy spectra
to higher energies, representative of non-thermal
emission.   Even in this optimistic case, it is difficult to
assess this lower energy spectrum owing to the source variability.
Simultaneous multi-wavelength observations are required.
However, life is expected to be more complex, and the intrinsic spectra
might consist of a sequence of humps, with spectral hardening and
softening in the observed energy window (this is the case, for instance,
in the proton-initiated cascade models, e.g. Mannheim 1998).
Moreover, the presence of strong infrared radiation fields within
the sources would lead to gamma ray cut-off energies well below
the cosmological ones (Donea \& Protheroe 2003). 
These effects would mostly place the measured pairs of cut-off energy and
redshift to the left of the expected FSR.
The obstacles can be overcome by collecting data of a large sample
of extragalactic gamma ray sources, and by matching the theoretical FSR
to the upper boundary curve for all entries in the FS diagram.

\subsection{Gamma-ray attenuation for low-redshift sources}
\begin{figure*}
  \centering
  \includegraphics[width=14cm]{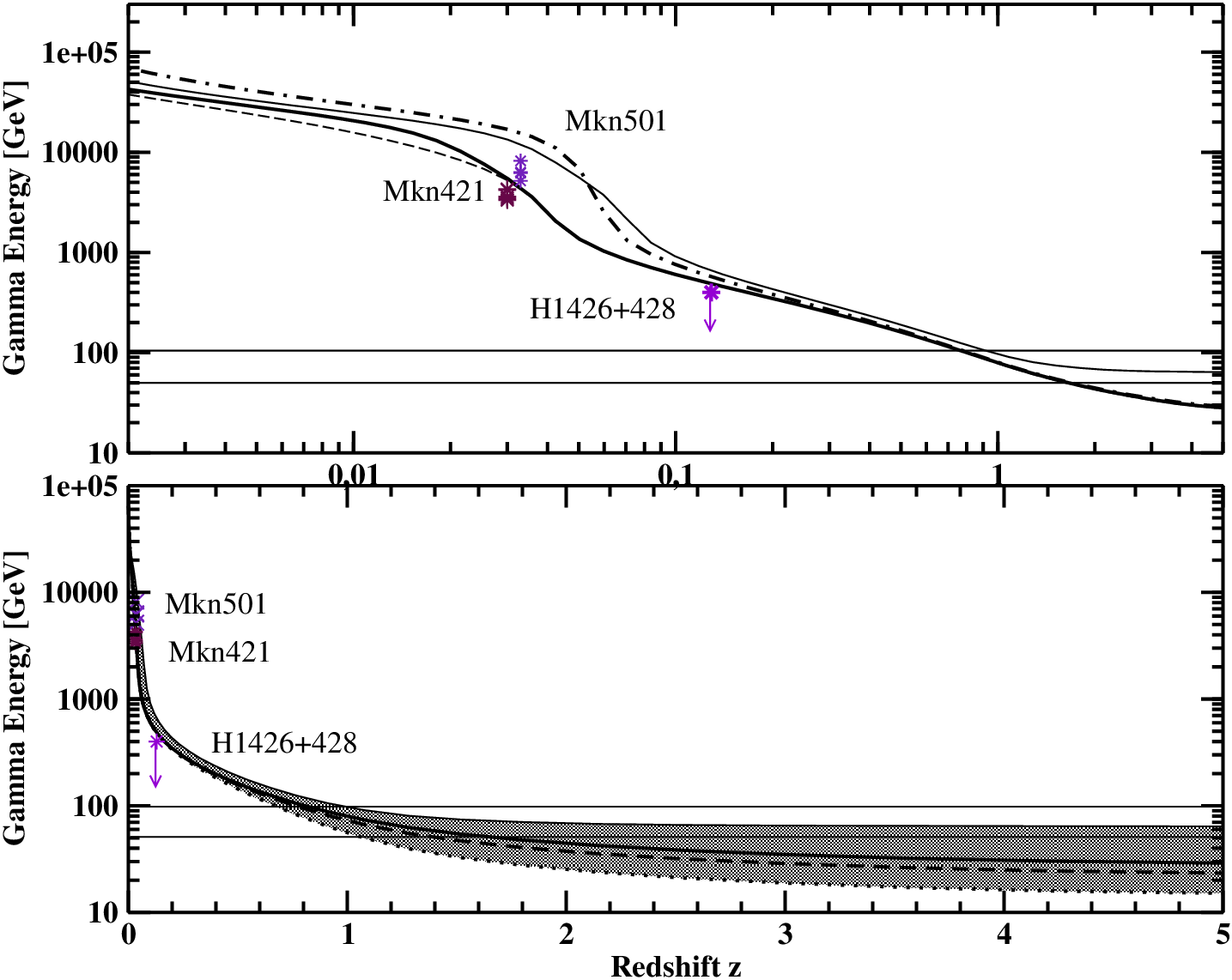}
  \caption{{\em Upper panel:} Fazio-Stecker relation with logarithmic redshift axis (best-fit model -- {\em thick solid line};
  low-IR model -- {\em dot-dashed line}; warm-dust model -- {\em thin dashed line} low-SFR model -- {\em thin solid line}). {\em Lower panel:} 
Fazio-Stecker relation with linear redshift axis showing the asymptotic far-zone
(best-fit model -- {\em thick solid line}; stellar-UV model -- {\em dashed line}; high-stellar-UV model -- 
{\em dotted line}; low-SFR model --
  {\em thin solid line)}. Also plotted are published cut-off energies of Mkn501, Mkn421, and H1426+428 
(for references, see text).  The horizontal lines at 50 GeV and 100 GeV represent
guide lines showing how the asymptotic branch of the Fazio-Stecker relation can be tapped by
lowering the detection threshold to below 50 GeV (e.g. using the MAGIC telescope).
}
  
  \label{fig4}
  \end{figure*}

In the redshift interval
$0.02<z<0.1$, three of the five used generic MRF parameter sets
show a clearly distinguishable behavior (see Figure~\ref{fig4}, upper panel).
The three sets are best-fit, low-IR, and low-SFR, whereas those sets in which the
UV fraction is varied naturally do not lead to measurable differences. 
The warm-dust model is only different from the best-fit model at redshifts 
smaller than 0.02 (dashed-line).
The published values
for the cut-off energies of the three nearby blazars seem to support the best-fit MRF
model which entails a strong far-infrared component in the present-day extragalactic
background.  No clear evidence for deviations from intrinsic power
law spectra emerges, a hardening of the intrinsic spectra would
allow for a yet stronger FIR component in the MRF model than adopted
in the best-fit model.  However, an increase by more than a factor of two 
would be inconsistent with the chemical abundances in the Universe
(Pei, Fall \& Hauser 1999).

\subsection{Gamma-ray attenuation for high-redshift sources}

At high redshifts, the Fazio-Stecker relation for
the high-IR and the low-IR models converge, since the optical
depth becomes independent of the density of infrared photons. Due to the large
distances, which increase the column depth for pair production,
the cut-off energies are generally at lower energies where the threshold condition
implies interactions with harder photons. Hence it follows that the
sensitivity on the diffuse UV radiation field is enhanced at high redshifts.
Considering UV radiation up to the level of the proximity data in the high-stellar-UV 
model, gamma-ray attenuation
for sources at redshifts $z>1$ increases considerably compared to the best-fit model, 
or the stellar-UV model.
The low star formation rate at high redshifts adopted in the low-SFR model 
would reduce the UV photon density, making the universe more transparent
for gamma-photons. However,  the large
drop of the star formation rate has been adopted only
to bracket the range of possibilities, while latest observations rather indicate 
a plateauish behavior for the 
star formation rate at high redshifts, or even a shallow upturn.

Note that the Fazio-Stecker relation in this redshift regime is also sensitive to 
cosmological parameters (Blanch \& Martinez 2001), 
which we have set to the current
$\Lambda$CDM cosmology values.

\section{Conclusions} 
Direct methods for measuring the MRF based on faint galaxy counts suffer from being limited to rather 
narrow wavelength ranges (introducing strong selection effects)
and from not being sensitive to a truly diffuse component of the MRF.  Complementary information
from an inherently independent method, such as measuring the
Fazio-Stecker relation is therefore of great diagnostic value to
determine the amount
of star formation occurring in optically-obscured infrared galaxies, 
to find other than stellar sources
of UV radiation at high redshifts, and to determine
the star formation rate at
high redshifts. 

A major source of uncertainty is the shape of the unabsorbed
gamma ray spectra, and how much of an observed cut off can be attributed
to cosmological absorption in the MRF.  In the Fazio-Stecker diagram,
sources with intrinsic gamma ray absorption would appear to lie
on the left side of the
FS relation.  This bias can be overcome statistically by observing
a large sample of
gamma ray sources over a broad range of redshifts to find
the upper bound in the FS diagram.
To probe the present-day infrared background at energies larger than
10$\mu$m using the FS diagram, cut-off energies from sources
at redshifts z$<$0.02 are needed. The only method to make predictions in this
part of the MRF is to look at single-source spectra at energies larger
than 10 TeV, dealing again with all uncertainties mentioned above.

Practically, this 
emphasizes the importance of very large
imaging air-Cherenkov telescops such as MAGIC (Corinta 2001, Martinez 2003) or the proposed 
ECO-1000 telescope (Martinez et al. 2003, Merck et al. 2003), 
which have the capability to discover
a large number of sources, and which achieve low-threshold energies.

The small range of the Fazio-Stecker relation which has been probed
with the current generation of IACTs indicates consistency
with the optical and infrared data on the MRF.
Lacking detections of high-redshift sources with 
cut-off energies below 100~GeV, no
conclusions about the UV part of the MRF can currently be drawn .

\begin{acknowledgement}
We thank Eli Dwek, Alberto Franceschini and Ranga-Ram Chary for providing their
MRF results and for discussions.
This research was gratefully supported by the BMB+f under grant 05AM9MGA.
\end{acknowledgement}

 \end{document}